\documentclass[10pt,a4paper,prl,twocolumn]{revtex4}
\usepackage{amsmath}
\usepackage{amsfonts}
\usepackage{amssymb}
\usepackage{graphicx}
\begin{document}

\title{Thermodynamics and Statistical Mechanics of dense granular media}
\author{Massimo Pica Ciamarra}\email[]{picaciamarra@na.infn.it}
\author{Antonio Coniglio}
\author{Mario Nicodemi}

\affiliation{Dipartimento di Scienze Fisiche, Universit\'a di
Napoli `Federico II', CNR-Coherentia, INFN, 80126 Napoli,
Italia.}
\homepage{http://smcs.na.infn.it}

\date{\today}

\begin{abstract}
By detailed Molecular Dynamics and Monte Carlo simulations 
we show that granular materials at rest can be described as thermodynamics systems.
First we show that granular packs can be characterized by few parameters, 
as much as fluids or solids. Then, in a second independent step, 
we demonstrate that these states can be described in terms of equilibrium distributions which coincide
with the Statistical Mechanics of powders first proposed by Edwards. 
We also derive the system equation of state as a function of the 
``configurational temperature'', its new intensive thermodynamic 
parameter. 
 \end{abstract}
\maketitle

Granular materials, such as sand or powders, 
for many respect are similar to fluids or solids~\cite{Jaeger}, 
even though in absence of external drive they rapidly 
come rest, due to strong dissipation and negligible thermal energy 
scales (they are non-thermal systems), in disordered states 
very similar to glasses~\cite{Nicodemi97,Liu1,Liu2,Silbert,KurchanMakse,Coniglio}. 
As standard thermodynamics is not applicable to describe them, it is 
natural to ask whether we can even refer to granular packs as ``states". 
The problem of finding the correct theoretical framework where to 
describe granular media is in fact of deep relevance to civil
engineering, geophysics and physics~\cite{Jaeger,Behringer,Nedderman,Siegfried}. 

Edwards proposed~\cite{Edwards} a thermodynamic description 
for static granular media, which was partially investigated by recent experiments~\cite{Knight95,Nowak97,Swinney05}.
These experiments have established that a granular system subject to a tapping dynamics, 
such as subsequent mechanical oscillations of the container, may loose memory of its 
initial state and reach a stationary state of volume fraction only dependent on the tapping 
intensity, a precondition for a statistical mechanics description of static granular material to be possible.
The study of out-of equilibrium (aging) slowly sheared granular assemblies is also useful for the validation
of the statistical mechanics of granular media~\cite{KurchanMakse}, but it is inherently restricted to a small range of very high volume fractions, where the system is jammed~\cite{potiguar}.

Here we give strong evidences supporting the existence of a thermodynamical and statistical mechanical description of granular media. First we demonstrate, via Molecular Dynamics (MD) simulations, that granular packs at rest are genuine thermodynamic states, as they are characterized by a small set of parameteres regardless of the procedure with which they are generated. Then we show, via Monte Carlo (MC) simulations, that these states can be described in terms of the 
equilibrium distribution proposed by Edwards. The coincidence between time averages (MD) and ensemble averages (MC) 
is a strong evidence in favour of the statistical mechanics approach to granular media. 
For details of materials and methods, see the supplementary materials available online~\cite{epaps}.

{\it MD simulations: time averages} --
\begin{figure}[t!!]
\begin{center}
\includegraphics*[scale = 0.38]{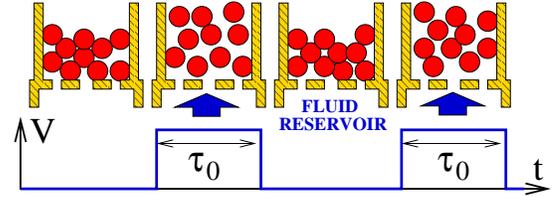}
\caption{\label{fig1} 
(color online) Schematic plot of the investigated system. A monodisperse 3D grain pack
of spheres immersed in a fluid is subject to a dynamics made of flow pulses of velocity $V$ 
and duration $\tau_0$. Before applying a pulse we wait for the pack to settle. 
The flow pulse dynamics is repeated up to reach stationary conditions (see Fig~\ref{fig2}).}
\end{center}
\end{figure}
We run Molecular Dynamics simulations of $N=1600$ monodisperse 
spherical grains of diameter $d = 1$cm and mass $m=1$g. Grains, under gravity, 
are confined in a box with a square basis of length $L = 10$cm 
(see Fig.~\ref{fig1}), with periodic boundary conditions in the horizontal directions. 
The bottom of the box is made of other immobile, randomly displaced,
grains (to prevent crystallization). Two grains in contact interact via a normal
and a tangential force. The former is given by the 
spring-dashpot model, while the latter is implemented
by keeping track of the elastic shear displacement throughout the
lifetime of a contact~\cite{Silbert1,modello}. The coefficient
of restitution is constant, $e= 0.8$. 
\begin{figure}[t!!]
\begin{center}
\includegraphics[scale=0.33]{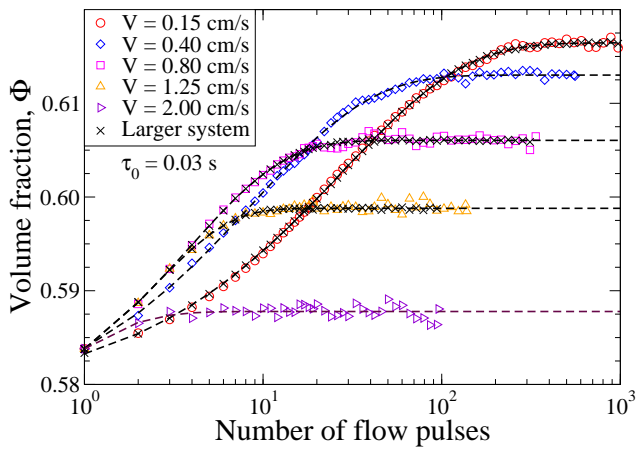}
\end{center}
\caption{\label{fig2} (color online) Compaction of systems subject to a tapping dynamics 
with $\tau_0 = 0.03$s and different values of the fluid velocity as indicated, averaged over $32$ runs.
Stars refers to simulations with $4$ times more particles, outlining the absence of finite size effects. 
Dashed lines are fits to a stretched exponential law, Eq.~\ref{eq-se}.}
\end{figure}

As in a recent experiment~\cite{Swinney05} the system is immersed in a fluid and, 
starting from a random configuration, it is subject to a dynamics made of a sequence of 
flow pulses where the fluid flows through the grains (see Fig.~\ref{fig1}).
In a single pulse the flow velocity, directed against gravity, is $V>0$ 
for a time $\tau_0$; then the fluid comes to rest. 
We model the fluid-grain interaction~\cite{King,Crowe} 
via a viscous force proportional to the fluid grain relative velocity:
${\bf F}_{fg}= -A({\bf v} - {\bf V})$
where ${\bf v}$ is the grain and ${\bf V}$ is the fluid velocity. The prefactor 
$A={\gamma}{(1-\Phi_l)^{-3.65}}$ is dependent on the local 
packing fraction, $\Phi_l$,
in a cube of side length $3d$ around the grain, and 
the constant~\cite{Crowe} is $\gamma = 1$ Ns/cm. 

During each pulse, grains are fluidized and then come to rest under
the effect of gravity. The tapping dynamics, therefore, allows for
the exploration of the phase space of the mechanically stable granular
packs. When the system is subject to such a tap dynamics,
it compactifies until it reaches a stationary state where its
properties do not depend on the dynamics history. 
Fig.~\ref{fig2} shows that the volume fraction of our system increases by following a stretched exponential 
low,
\begin{equation}
\label{eq-se}
\Phi(t) = \Phi_\infty - (\Phi_\infty - \Phi_0) \exp\left(-(t/\tau)^c\right),
\end{equation}
in agreement with the experiment by P. Philippe {\it et al.}~\cite{Philippe}.
The relaxation time diverges as the tapping intensity decreases, indicating 
the presence of a glassy like behavior which will be discussed elsewhere~\cite{futuro}.
As the thermodynamics approach to granular media aims to describe stationary states,
all of measures shown below (averaged over $32$ runs) are recorded after the application
of a long sequence of flow pulses, when the system is at stationarity.

\begin{figure}[t!!]
\begin{center}
\includegraphics*[scale = 0.35]{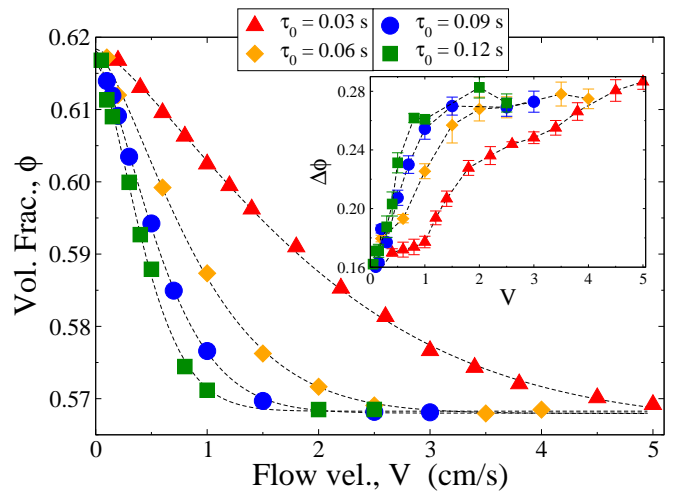}
\end{center}
\caption{\label{fig3} 
(color online) The main panel shows the volume fraction $\phi$ attained
at stationarity (see Fig.~\ref{fig2}) measured when the pack is
settled, as a function of $V$, for the shown values of $\tau_0$: 
the stronger the pulses the fluffier the pack. The inset shows
the dependence of the volume fraction fluctuations on $V$ 
and $\tau_0$.}
\end{figure}

We plot in Fig.~\ref{fig3} the stationary values of the volume
fraction, $\phi(V,\tau_0)$ (measured in the bulk of the system), and its fluctuations, 
$\Delta \phi(V,\tau_0)$, recorded after a sequence of such flow pulses 
of duration $\tau_0$ and velocity $V$.
$\Delta \phi$ is by definition the standard deviation of $\phi$ around its average value at stationarity. 
Actually, the volume fraction probability distributions is Gaussian~\cite{Swinney05,epaps}.
$\phi$ decreases with $V$ and with $\tau_0$: the
stronger the pulse, i.e., the larger $V$ or $\tau_0$, the fluffier the
pack settled after it. Similarly, $\Delta \phi$ increases with $V$ and $\tau_0$. 

\begin{figure}
\includegraphics{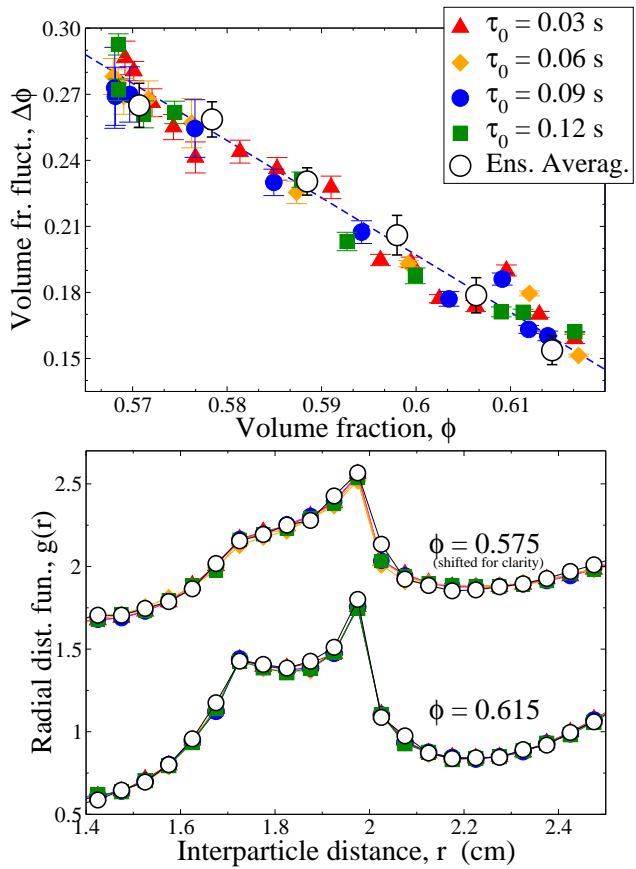}
\caption{\label{fig4}(color online) {\bf Upper panel} 
The $\Delta \phi$ data of Fig.~\ref{fig3} (same symbols are
used) are here plotted as a function of $\phi$: they collapse on a 
single master curve, showing that $\Delta \phi$ is in a one-to-one 
correspondence with $\phi$, irrespectively of the dynamical protocol
used to arrive at $\phi$. The dotted line is a linear best fit. 
A similarly good collapse is also found ({\bf lower panel})  when
we plot the radial distribution function $g(r)$ for packs having the 
same volume fraction (the same packs of Fig.~\ref{fig3} are shown with the same symbols). 
In particular, we show $g(r)$ for $\phi=0.615$ and $\phi=0.575$ (shifted for clarity). 
The empty circles in both panels are the corresponding 
ensemble averages independently calculated from Eq.~\ref{pr}: within
numerical errors, they scale on the same curves, pointing out this 
Statistical Mechanics measure is an excellent approximation of the
time averages over the pulse dynamics. 
}
\end{figure}

Even though, $\phi$ depends on both the parameters of the dynamics, $V$ and
$\tau_0$, we show now that such stationary states are indeed genuine 
``thermodynamic states'', i.e., they can be described, in this system, 
by one macroscopic parameter. 
Actually, the upper panel of Fig.~\ref{fig4} 
shows that when $\Delta \phi$ is parametrically
plotted as function of $\phi$, the scattered data of Fig.~\ref{fig3} 
collapse, within numerical approximation, onto a single master
function. This is a clear indication that $\Delta \phi$ and $\phi$ are
in a one-to-one correspondence, no matter how the state with packing
fraction $\phi$ is attained. Our claim is that such a property should be found
for any macroscopic observable of the system: we checked some of them, 
including the energy and its fluctuations and the coordination number of grains. Actually,
in Fig.~\ref{fig4} lower panel we show that the whole radial distribution function
$g(r)$ of a pack is characterized only by its corresponding value of $\phi$, i.e., states attained with different dynamical protocols ($V,\tau_0$), but having the same $\phi$, have the same $g(r)$. 
From these results we derive our first conclusion: at stationarity, 
we can describe the pack with only one parameter, e.g., $\phi$,
independently of the dynamical protocol. Such a parameter
characterizes, thus, the ``thermodynamics state'' of the system.

{\it MC simulations: ensamble averages} --
Our second important step is to identify the
correct Statistical Mechanics distribution for these states. 
Under a very strong assumption 
(discussed for instance in~\cite{Coniglio,Brey00,Edwards,Dean01,Coniglio01,Barrat00,Fierro02,Ono02,DeSmedt03,Tarjus04,Richard,epaps}), 
Edwards proposed to use for the grains of a 
powder the standard machinery of Statistical Mechanics. He suggested, 
however, to consider a reduced configurational space: the system at rest 
(i.e., not in its ``fluidized'' regime) is described by a flat ensemble average restricted to 
its blocked configurations (i.e., its mechanically stable microstates). 
Under these hypotheses~\cite{Coniglio01,Fierro02,epaps}
the canonical ensemble probability, $P_r$, to find the blocked
microstate $r$, of energy $E_r$, is: 
\begin{eqnarray}
P_r\propto e^{-\beta_{conf} E_r},
\label{pr}
\end{eqnarray}
where the inverse of $\beta_{conf}$ is the conjugate parameter of the
energy, called {\tt configurational temperature}, $T_{conf}$.

In order to check whether such a Statistical Mechanics scenario
applies, we compared ensemble averages over the distribution
of Eq.~\ref{pr} with those over the flow tap dynamics. 
For instance, the average value of $\phi$ over the distribution of
Eq.~\ref{pr} is
\begin{equation} 
\langle \phi\rangle(T_{conf})=
\frac{\sum_r\phi_r\exp(- E_r/T_{conf})}{\sum_r\exp(- E_r/T_{conf})},
\end{equation} 
where the sum runs over all blocked microstates, and $\phi_r$ is
the volume fraction of microstate $r$. We evaluated these ensemble averages by use of a
Monte Carlo method which is an extension of that introduced in Ref.~\cite{KurchanMakse}
to the frictional case~\cite{epaps} (frictional forces are essential to assure the
stability of granular packs with small volume fraction).
Fig.~\ref{fig4} shows, as empty circles, the functions $\langle \Delta \phi\rangle(\langle \phi\rangle)$ 
(resp. $\langle g(r) \rangle(\langle\phi\rangle)$)
in the upper (resp. lower) panel.
These ensemble averages
collapse, to a very good approximation, on the same
master function of the time averaged data from the flow-pulse dynamics 
discussed before (notice that there are no adjustable parameters). 
Thus, the present Statistical Mechanics description appears to hold, 
up to the current numerical accuracy, at least as a first very good
approximation. 
Interestingly, the off-equilibrium dynamical effective temperature defined
at high volume fractions from dynamical fluctuation-dissipation relations~\cite{KurchanMakse}
appears to coincide with the configurational temperature derived here at stationarity~\cite{epaps}.

\begin{figure}
\includegraphics{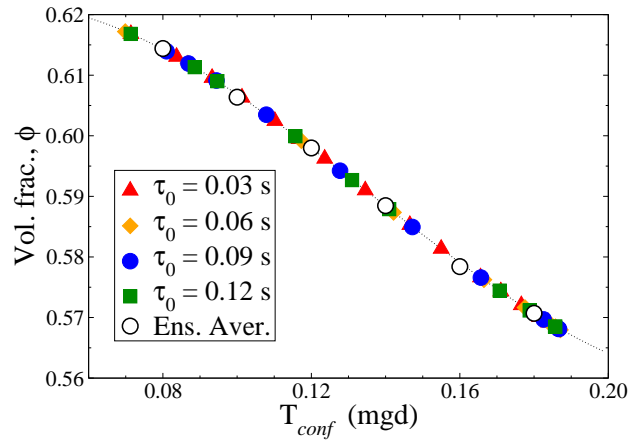}
\caption{\label{eq_state} (color online) The equation of state of granular materials
at rest, showing the volume fraction, $\phi$, as a function of the
configurational temperature, $T_{conf}$. 
Ensamble averages are obtained via the Monte Carlo procedure, while
time averages are obtained from the data of Fig.~\ref{fig3} via the
use of the static fluctuation dissipation relation~\cite{epaps}.
}
\end{figure}

The function $\phi(T_{conf})$, derived from the above 
ensemble averages, is the system {\tt equation of state}. 
We plot $\phi(T_{conf})$ in Fig.~\ref{eq_state}, where the data from MD 
simulations are included too by using the data collapse from 
Fig.~\ref{fig4} and integration of the static
fluctuation-dissipation relations~\cite{Knight95,Coniglio01,Fierro02,Swinney05,epaps}.

{\it Conclusions} --
Summarizing, in the present MD simulations of a non-thermal
monodisperse granular system under flow pulses, we find that 
the stationary configurations of the system can be 
fully described by only one parameter, e.g., $\phi$, and can be, thus, 
considered genuine ``thermodynamic states''. Within our numerical
accuracy, we also showed that a Statistical Mechanics based on the
distribution of Eq.~\ref{pr} is grounded to describe these
``states''. We could derive as well the equation of state,
$\phi(T_{conf})$, of the system. 


These evidences strongly support the existence of
a fundamental theory of dense
granular media and address, thus, a variety of important issues
in the next future, such as response functions in a granular system,
mixing/segregation phenomena, the nature of their jamming
transition and phase diagram~\cite{Coniglio,Liu1,Richard,Corwin}.

\newpage
\begin{widetext}
\oddsidemargin  1cm
\textwidth 14cm
\textheight 22cm
\pagestyle{empty}
\noindent
\begin{center}
\vspace{1cm}
\Large{Thermodynamics and Statistical Mechanics\\ of dense granular media}
\\
\Large{Supplementary Material}
\end{center} 
\vspace{2cm}
\begin{large}
Contents:\\
\begin{enumerate}
\item The Statistical Mechanics approach to granular media.
\item Configurational temperature and dynamical temperature.
\item Monte Carlo method.
\end{enumerate} 
\end{large}
\newpage
\noindent {\Large \bf Volume fraction fluctuations in the stationary state}
\begin{figure}[h!!!]
\begin{center}
\includegraphics*[scale=0.45]{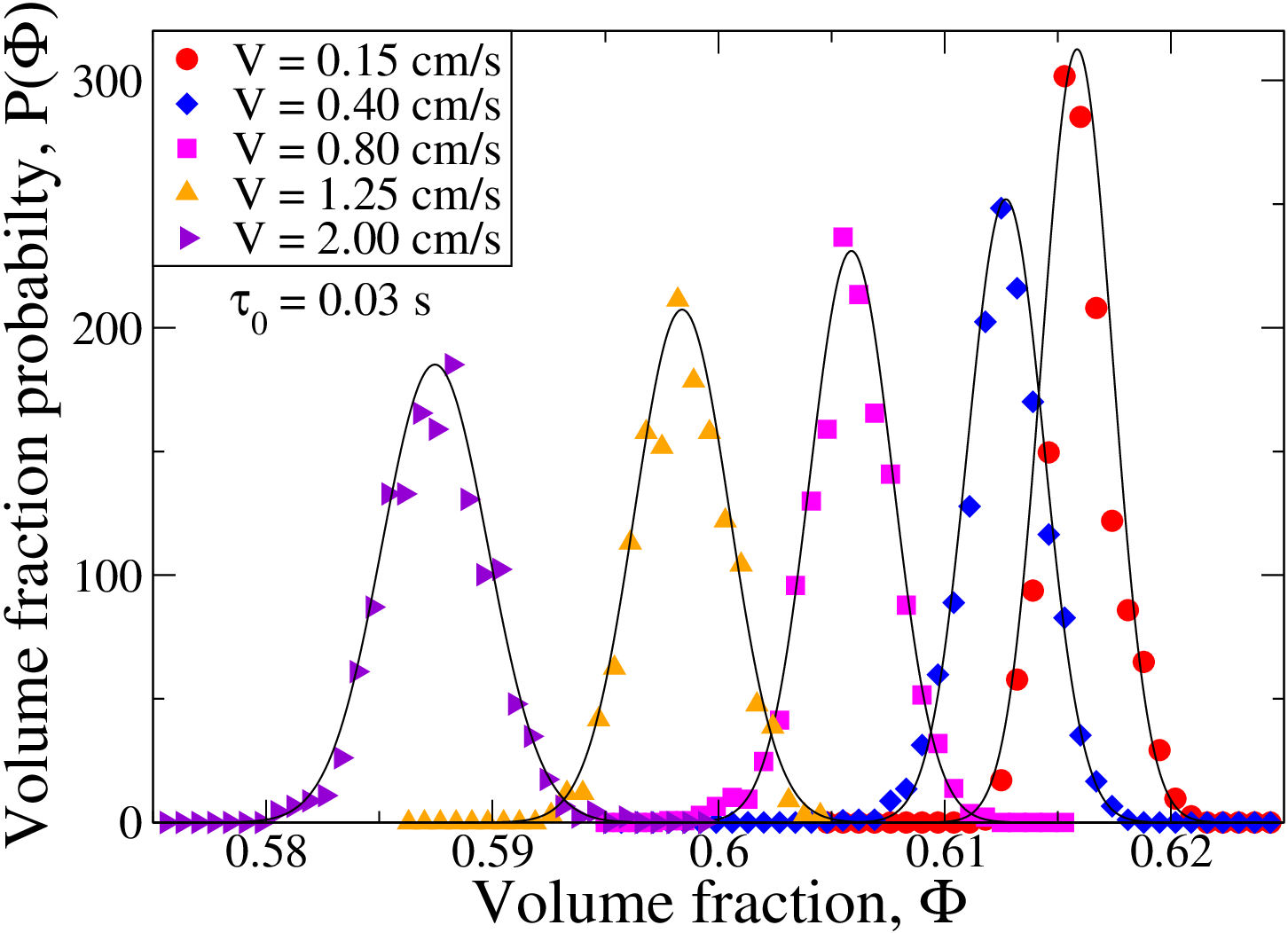} 
\end{center}
\end{figure}

\noindent {\bf Figure A} Probability distributions of the volume fraction $\Phi$ in the steady state reached by the system subject to the flow pulses dynamics. The data, obtained by averaging over the stationary dynamics in $32$ different runs starting from random initial conditions, are Gaussian distributed (plain lines). The standard deviations of these distributions are the volume fraction fluctuations, $\Delta \Phi$, plotted in Fig. 3.

\newpage

\noindent {\Large \bf Statistical Mechanics approach to granular media}\\
In the Statistical Mechanics of powders introduced by S. Edwards (S.F. Edwards and R.S.B. Oakeshott, Physica A {\bf 157}, 1080, 1989) it is postulated that the system at rest
can be described by suitable ensemble averages over its ``mechanically stable'' states. Edwards proposed
a method to individuate the probability, $P_r$, to find the system in its mechanically stable state $r$,
under the assumption that these mechanically stable states have the same a priori probability to occur.
This is the simplest assumption one can imagine: it is the assumption of standard Statistical Mechanics (equiprobability of microstates) with the additional constraint of mechanical stability.
A possible approach to find $P_r$ is as follow (A. Fierro {\it et al.}, Europhys. Lett. {\bf 59}, 642, 2002; Phys. Rev. E {\bf 66}, 061301, 2002; Europhys. Lett. {\bf 60} 684, 2002). $P_r$ is
obtained as the maximum of the entropy,
\begin{equation}
S = -\sum_r P_r \log P_r
\nonumber
\end{equation}
with the macroscopic constraint, in the case of the canonical ensemble, of fixed system energy $E = \sum P_r E_r$ (here $E_r$ is the energy of the mechanically stable microstate $r$). For a granular medium at rest in the gravity field, $E_r$ is the sum of the gravitational energy, and of the interaction energy between grains:
\begin{equation}
E_r = \sum_i mgz_i + \sum_{i\neq j} V_{ij},
\label{eq-energy}
\end{equation}
where $z_i$ is the height of grain $i$ (with mass $m$) and $V_{ij}$ is the interaction (elastic energy) between grain $i$ and $j$ in the microstate $r$: $V_{ij} = 0$ if grains $i$ and $j$ are not in contact, otherwise 
\begin{equation}
V_{ij} = \frac{1}{2} k_n |\vec \delta_{ij}|^2 + \frac{1}{2} k_t |\vec u_{ij}|^2,
\end{equation}
where $\vec \delta_{ij}$ is the overlap between grain $i$ and $j$, and $\vec u_{ij}$ is their shear displacement. The interaction energy $V_{ij}$ is derived by the normal and the tangential forces acting on the contacting grains (we use the model L3 described in L.E. Silbert {\it et al.}, Phys Rev E {\bf 64}, 051302, 2001): $\vec f^n = k_n \vec \delta_{ij}, \vec f^t = k_t\vec u_{ij}$.

From Edwards' hypothesis, in analogy to the Gibbs result, you derive that:
\begin{equation}
P_r = Z^{-1} \exp\left(-\beta_{conf}E_r\right)
\nonumber
\end{equation}
where $\beta_{conf}$ is a Lagrange multiplier, called inverse configurational temperature, enforcing the above constraint on the energy:
\begin{equation}
\beta_{conf} = \frac{\partial S_{conf}}{\partial E}; \;\;\; S_{conf} = \ln \Omega_{\rm stable}(E).		
\nonumber
\end{equation}
The configurational temperature is $T_{conf} = \beta_{conf}^{-1}$. Here $\Omega_{\rm stable}(E)$ is the number of mechanically stable states with energy $E$. $Z$ is fixed by the normalization condition $\sum_r P_r = 1$, where the sum is restricted to $\Omega_{\rm stable}$, i.e., to the mechanically stable states.\\
\\
\noindent We note here that for stiff grains the elastic energy $\sum_{i\neq j} V_{ij}$
is much smaller than the gravitational energy $\sum_i mgz_i$ (the elastic energy is
strictly zero in hard sphere systems). Accordingly $E_r$ is to a good approximation
proportional to the gravitational energy, i.e. to the volume of the system, 
as originally suggested by Edwards.

\newpage
\noindent {\Large \bf Configurational temperature and dynamical temperature}\\
We describe here the two different definitions of temperature for 
granular systems mentioned in our paper. 
\\
\\
\noindent {\bf Configurational temperature}\\ 
As discussed above, the configurational temperature is the inverse of the derivative of the
configurational entropy with respect to the energy. It is therefore an
`equilibrium' temperature, defined for a granular system at stationarity under a
given dynamics allowing the exploration of mechanically stable states.
\\
\\
\noindent In our main text we have validated Statistical Mechanics
approaches to powders 
by showing that, within numerical errors, the granular packs we
consider are characterized, at stationarity,  
by a single parameter regardless of the dynamical procedure with which they were prepared 
(they are in a thermodynamic `state'), and that time averages coincide with 
ensemble averages over the distribution detailed above. 
This result justifies the use of Edwards `equilibrium' partition function
to make analytical calculations, and 
particularly to derive the following equilibrium
fluctuation-dissipation relation (Coniglio and Nicodemi Physica A 
{\bf 296}, 451 (2001)),
\[\beta_{conf}(E) = \beta_{conf}(E_0) - \int_{E_0}^{E} \frac{dE}{\Delta E^2}\]
which relates the energy and its fluctuations. 
This relation can be exploited to experimentally measure the configurational temperature
(Nowak {\it et al.}, Powder. Tech. {\bf 94}, 79, 1997; Knight {\it et al.}, Phys. Rev. E {\bf 57}, 1971, 1998;
Schr\"oeter and Swinney, Phys. Rev. E {\bf 71}, 030301(R), 2005).
\\
\\
\noindent {\bf Dynamical temperature}\\
In thermal glassy systems far from stationarity, dynamical
off-equilibrium fluctuation-dissipation relations hold.
Particularly, in the aging dynamics of mean-field glassy models in
contact with a very small bath temperature,
$T_{\rm bath}$, generalized out-of equilibrium fluctuation-dissipation
relations were discovered where the role of the external temperature
is played by a `dynamical temperature' $T_{\rm dyn} \neq T_{\rm
  bath}$, equal for all slow modes (L.F. Cugliandolo and J. Kurchan,
Phys. Rev. Lett. {\bf 71}, 173, 1993). This scenario was later extended to 
aging granular materials (for a review see P. Richard {\it et al.},
Nat. Mat. {\bf 4}, 121, 2005, and references therein),
were the dynamical temperature is defined and measured via a
dynamical, off-equilibrium fluctuation dissipation relation,
\begin{equation}
\nonumber
\langle (r(t)-r(t_w))^2 \rangle = T_{\rm dyn}\frac{\delta\langle r(t)-r(t_w) \rangle}{\delta f}
\end{equation}
where $r$ is the position of a grain, $f$ a constant perturbing field,
and $t_w$ the `waiting' time.
\\
\noindent 
As the dynamical temperature is defined far from stationarity and,
conversely, the configurational temperature at stationarity, 
Makse and Kurchan have
shown that in granular packs, at very high density, the two are
equal within numerical errors (Nature {\bf 415}, 614, 2002). 
At low volume fraction, on the contrary,
the dynamical temperature appears to be no longer defined, as the granular system is no longer jammed
and flows as soon as an external perturbation is applied, as shown by 
Potiguar and Makse (European Physical Journal E 
{\bf 19}, 171, 2006). 

\newpage
\noindent {\Large \bf Monte Carlo method to test the Statistical Mechanics approach to granular materials at rest}
\newline
\newline
In order to test the Statistical Mechanics approach to granular media one has to compare
time averaged data, obtained as explained in the text, with ensemble averaged data obtained by sampling Edwards distribution, eq.(2) in the text,
\begin{equation}
P_r = Z^{-1}\exp(-\beta_{conf}E_r).
\label{eq-edwards}
\end{equation}
where $E_r$ is the energy of the state $r$,
\begin{equation}
E_r = \sum_i mgz_i + \sum_{i\neq j} V_{ij}.
\label{eq-energy1}
\end{equation}

As the Statistical Mechanics approach to granular media deals with mechanically stable states, the phase space
of interest (over which eq.~(\ref{eq-edwards}) must be sampled) is not the usual phase space, $\Omega_{\rm tot}\{\vec r, \vec v, \vec \omega \}$
(here $\vec r$ is the $3N$ vector of grains c.o.m. positions, $\vec v$ their 
velocities, and $\vec \omega$ their angular velocities). 
Instead it is the subset $\Omega_{\rm stable} \subset \Omega_{\rm tot}\{\vec r, 0, 0 \}$ of all states $r$ where the forces and the torques acting on each single grain sum to zero, and grains neither translate nor rotate.
\newline
Since the states to be considered are so highly constrained it is difficult to sample the distribution of eq.~(\ref{eq-edwards}) via a standard Monte Carlo procedure. For instance, if a state $r$ is stable, then there is little chance to transform it in a new mechanically stable state $r'$ via the displacement of a single particle. Introducing many particles Monte Carlo moves is also useless as the probability of selecting a collective move that transform a mechanically stable state into a new mechanically stable state is practically zero.
\newline
A Monte Carlo (MC) method to explore $\Omega_{\rm stable}$ was proposed by 
H.A. Makse and J. Kurchan, Nature {\bf 415} 614 (2002), which uses the following computational trick. The MC algorithm explores the usual phase space, $\Omega_{\rm tot}$, but in this phase space one introduces an auxiliary energy $E_{aux}$ which measures the degree of `mechanical instability' of a pack in a microstate $r$. In the present case, we have defined:
\begin{equation}
E_{aux} = \sum_{i=1}^N |m\vec{g}d + \sum_{j\neq i} (\vec f_{ij}^n + \vec f_{ij}^t)| + \sum_{i=1}^N |\vec{T}_i|,
\label{eq-eaux}
\end{equation}
where $\vec{T}_i$ is the total torque acting on grain $i$. 
The first term of the above equation enforce the balance of forces on each single grain, and the second the balance of torques. 
Different expressions could be used for $E_{aux}$, the important point being that $E_{aux}(r) \ge 0$ $\forall r$,
and that $E_{aux}(r) = 0$ if and only if the state $r$ is mechanically stable. 

Then one samples via a standard Monte Carlo procedure the distribution
\begin{equation}
\overline{P}_r = \overline{Z}^{-1}\exp(-\beta_{conf}E_r-\beta_{aux}E_{aux}(r)),
\label{eq-prob-aux}
\end{equation}
where $T_{aux}=\beta_{aux}^{-1}$ is the so-called `auxiliary temperature' which controls the equilibrium value of the auxiliary energy.
By definition of $E_{aux}$, in the limit $T_{aux}\to 0$ we have:
\begin{equation}
P_r = \lim_{T_{aux}\to 0} \overline{P}_r,\;\;\;
\lim_{T_{aux}\to 0} E_{aux} = 0.
\end{equation}
Therefore in the limit $T_{aux} \to 0$ we sample the distribution of mechanically stable states ($E_{aux} = 0$) with probability $P_r$, as desired.
This is precisely the limit which is considered in by Makse and Kurchan method. 

In our simulations we start by sampling eq.~(\ref{eq-prob-aux}) in the phase space of all granular packs at the desired value of $T_{conf}$ and at $T_{aux} > 0$ via a Monte Carlo procedure (see below). Then we slowly decrease the auxiliary temperature (carefully checking for thermalization) until $T_{aux} = 0$. By repeating this procedure severals times we generate several packs (a total of $172$) at the desired configurational temperature. These packs are then used to compute ensemble averages relative to the chosen value of $T_{conf}$.\\
\\
\noindent {\large \bf Implementation of the Monte Carlo method}\\
\noindent In our definition of mechanical stability (eq.~(\ref{eq-eaux})) we also include tangential forces 
(neglected in H.A. Makse and J. Kurchan, Nature {\bf 415} 614, 2002). Tangential forces model friction,
which is essential for our purposes since, under gravity, frictionless stable packs are only found for high volume fractions.

In MD simulations friction has important effects on the dynamics of the system: the frictional force between two grains depends on their shear displacement. In the Monte Carlo algorithm, however, it is convenient to consider the frictional force between two grains (that is their shear displacement) as an independent variable. This leads to the definition of two kind of MC moves: standard single particle displacement, and variation of the tangential shear displacement. The idea of separating geometrical properties of the pack (particle displacement) from the tangential forces (tangential shear variation) has been exploited previously in the literature, even though in a different contest and with simpler models (as for example in T.Unger, J.Kert\'esz, and D. E. Wolf, Phys. Rev. Lett. {\bf 94}, 178001 (2005); S. McNamara, R. Garcma-Rojo, and H. Herrmann, Phys. Rev. E  {\bf 72}, 021304 (2005)). Below we discuss briefly the two moves.\\ 
\\
\noindent {\bf Single-particle displacement}\\
\noindent One selects a particle $n$ in position $\vec r_n$ and a random displacement vector $\vec \Delta$, with $|\vec \Delta| < \lambda$ and $\lambda$ dynamically varied in order to obtain an acceptance probability $p_{\rm acc} = 0.5$. 
The displacement of particle $n$ from position $\vec r_n$ to position $\vec r_n + \vec \Delta$ induces a variation $\Delta E_r$ of the energy (eq.~(\ref{eq-energy})), and of the auxiliary energy, $\Delta E_{aux}$ (eq.~(\ref{eq-eaux})), of the system. $\Delta E_{aux}$ is due to:
\begin{enumerate}
\item changes of the overlaps $\vec \delta_{ni}$ between the displaced particle $n$ and a particle $i$,
and therefore of the normal force $\vec f_{ni}^{n}$. In particular contacts may disappear (after the displacement $| \vec \delta_{ni}  < 0|)$, or appear (before the displacement $|\vec \delta_{ni}  < 0|$, after the displacement $|\vec \delta_{ni}  > 0|)$.
\item variations of the shear displacements $\vec u_{ni}$, and therefore of the tangential force $\vec f_{ni}^{t}$. If particles $i$ and $n$ are in contact both before and after the displacement of particles $n$, we consider particle $n$ as sliding over particle $i$, inducing a variation of the shear displacement $\vec u_{ni}$. If the displacement of particle $n$ creates (destroys) a contact, then the shear displacement $\vec u_{ni}$ is created (destroyed) accordingly.
\end{enumerate} 
The shear displacement is rescaled by a factor $\mu |\vec f^n_{ij}| / k_t |\vec u_{ni}|$ if the Coulomb condition is violated. A move of this kind is accepted with probability $\exp(-\beta_{conf}\Delta E_r - \beta_{aux}\Delta E_{aux})$.\\
\\
\noindent {\bf Shear-displacements variation}\\
\noindent In this MC move one varies the tangential force between two contacting grains, 
respecting the Coulomb criterion. Two particles $n$ and $m$ are selected. If they are in contact one varies their shear displacement $\vec u_{nm}$ of a random amount $\vec \Delta_t$, with $|\vec \Delta_t| < \sigma$ and $\sigma$ dynamically varied in order to obtain an acceptance probability $p_{\rm acc} = 0.5$. $\Delta_t$ is actually chosen in such a way that $\vec u_{nm}+\Delta_t$ (and therefore the tangential force) lies in the plane tangent to both grains in their point of contact. Before varying $\vec u_{nm}$ by $\vec \Delta_t$ this latter is rescaled in order to satisfy the Coulomb criterion, if necessary.

The tangential move leads to a variation of both the energy of the system, $E_r$ (as the elastic energy in tangential interaction between the two particles varies), and of its auxiliary energy, $E_{aux}$. 
As before, the move is accepted with probability $\exp(-\beta_{conf}\Delta E_r - \beta_{aux}\Delta E_{aux})$.\\
\\
\noindent {\bf Generation of mechanically stable states with the MC algorithm}
\begin{figure}[h!!!]
\begin{center}
\includegraphics*[scale=0.45]{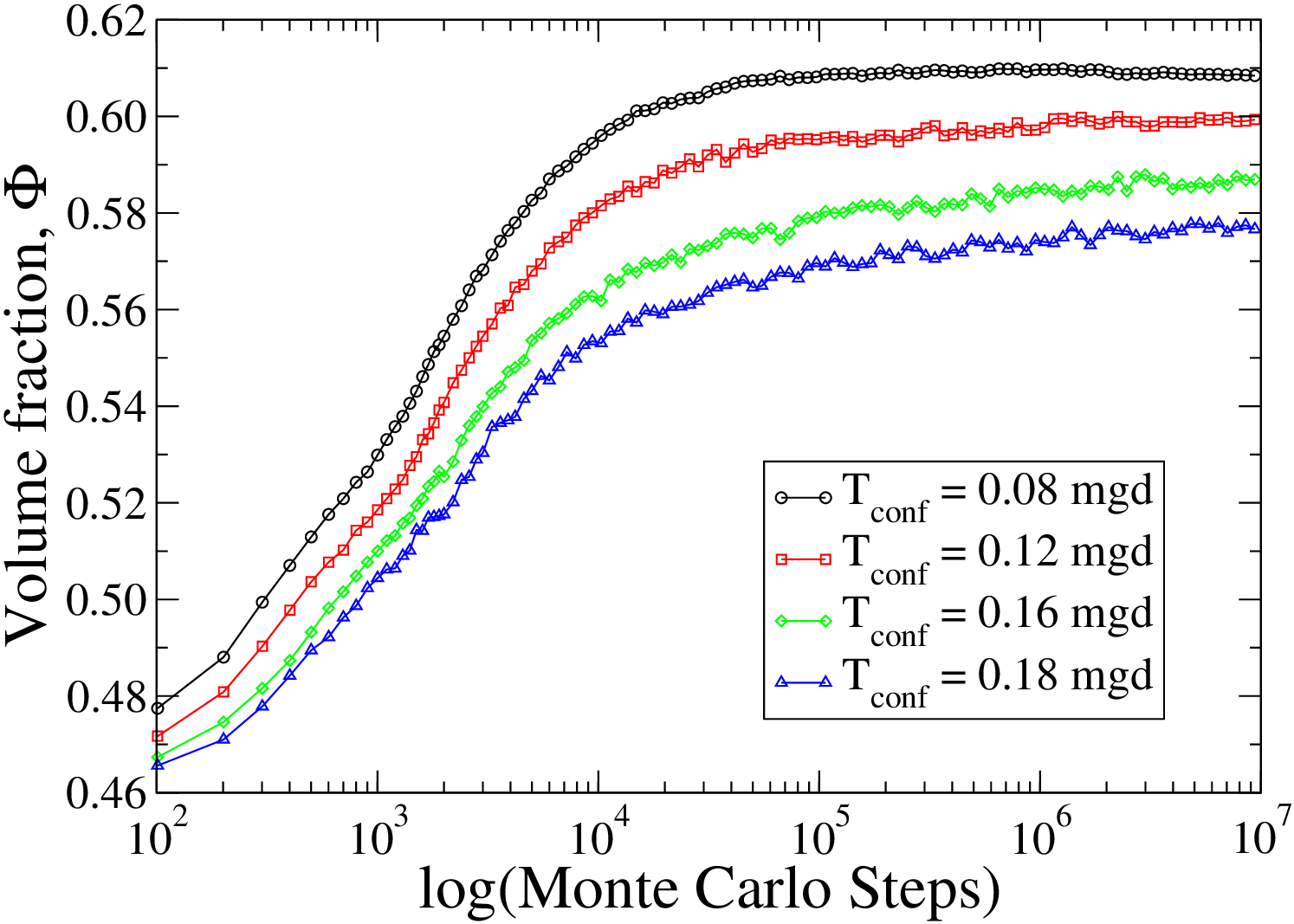} \\
\vspace{0.5cm}
\includegraphics*[scale=0.45]{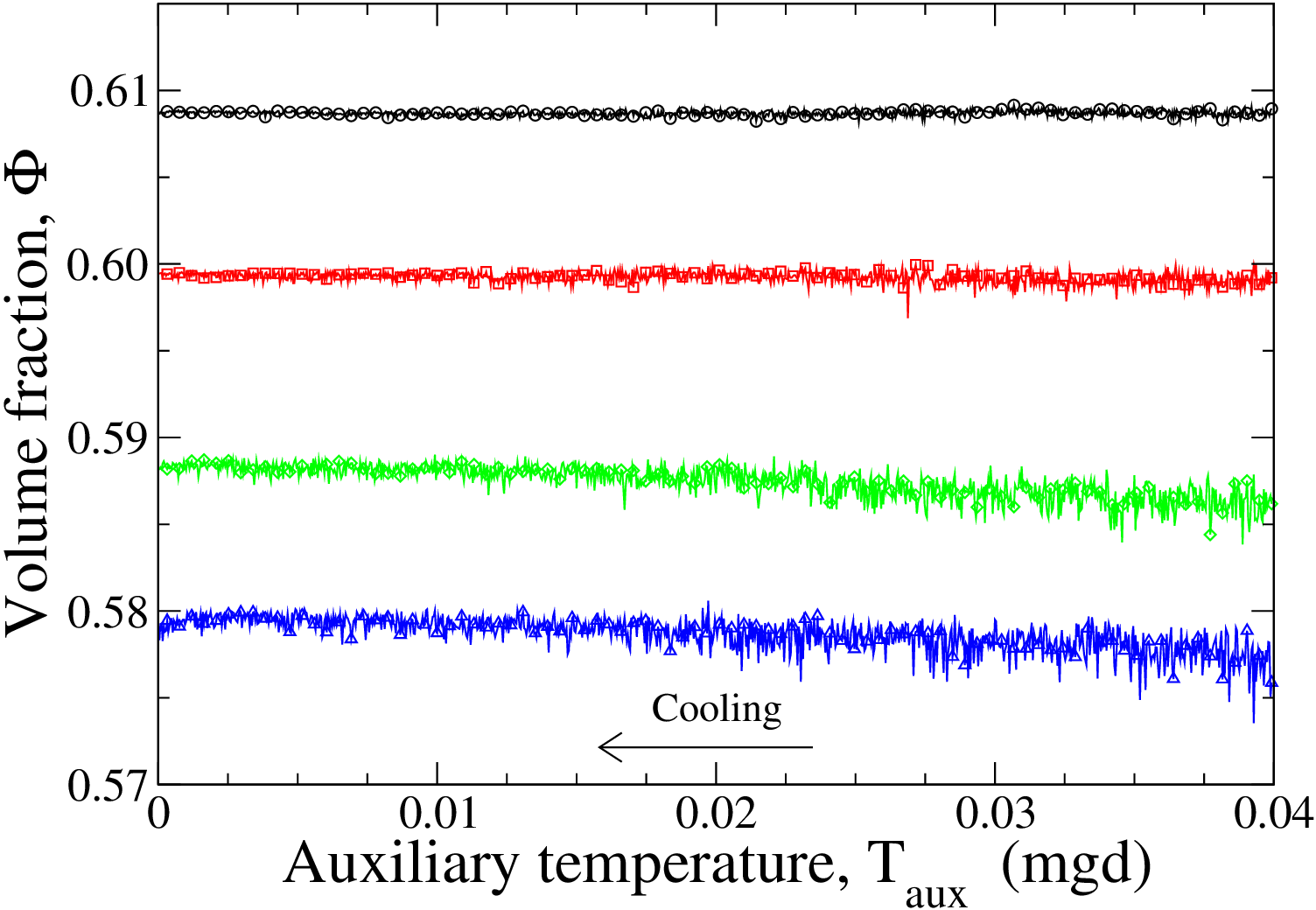} 
\end{center}
\end{figure}

\noindent Here we show how the volume fraction of the system 
evolves during the MC procedure. First we thermalize the system
at the desired value of $T_{conf}$ and at $T_{aux} = 0.04$ mgd (upper panel). Different values of $T_{aux}$ may be convenient in different runs. Then we slowly decrease the auxiliary temperature until the system reaches a mechanically stable state (lower panel). We check that the same results are obtained with slower cooling rates. 

\end{widetext}

\begin{thebibliography}{40}
\bibitem{modello}
We use the model `L3' described in detail in Ref.~\cite{Silbert1}.
\end{thebibliography}
\end{document}